# Faint active galactic nuclei supplied $31 - 75\%$ of hydrogen-ionizing photons at $z > 5$

Mainak Singha ,[1, 2, 3] Sangeeta Malhotra,[1] and James Ely Rhoads[1]

[1]*Astrophysics Science Division, NASA, Goddard Space Flight Center, Greenbelt, MD 20771, USA*
[2]*Department of Physics, The Catholic University of America, Washington, DC 20064, USA*
[3]*Center for Research and Exploration in Space Science and Technology, NASA, Goddard Space Flight Center, Greenbelt, MD 20771, USA*

## ABSTRACT

The origin of the ionizing photons that completed hydrogen reionization remains debated. Using recent *JWST* and ground-based surveys at $4.5 \leq z \leq 6.5$, we construct a unified rest–UV AGN luminosity function that separates unobscured Type I and obscured Type II populations, and show that "little red dots" and X–ray selected sources are magnitude–filtered subsets of Type I with mixture fraction $\eta = 0.10 \pm 0.02$. We anchor the Lyman–continuum (LyC) escape fraction to outflow incidence and geometric clearing rather than assuming quasar-like values for all classes, and propagate uncertainties through a joint fit. Integrating over $-27 < M_{\rm UV} < -17$, AGN inject $\dot{N}_{\rm ion}^{\rm AGN} = (3.77^{+1.08}_{-0.95}) \times 10^{51}\ {\rm s^{-1}\,Mpc^{-3}}$, nearly twice earlier estimates and comparable to the Ly$\alpha$-inferred requirement at $z \simeq 6$. When combined with the *JWST* galaxy UV luminosity function and a harder stellar ionizing efficiency of $\Psi_{\rm ion} = \log_{10}\xi_{\rm ion} = 25.7$, AGN contribute 31–75% of the total ionizing photons for representative escape fractions $f_{\rm esc}^{\rm gal} = 0.03$–$0.20$. The resulting hydrogen photoionization rate, $\Gamma_{\rm HI} \simeq (0.5\text{–}2) \times 10^{-12}\ {\rm s^{-1}}$ at $z \simeq 5$–6, lies squarely within the Ly$\alpha$ forest constraints once mean free paths and IGM clumpiness are accounted for, remaining consistent for combined AGN–galaxy models up to $f_{\rm esc}^{\rm gal} \leq 5\%$. These results suggest that AGN and galaxies jointly sustained the ionizing background during the final stages of reionization, with AGN remaining a major—but not exclusive—contributor.

*Keywords:* galaxies: AGN

## 1. INTRODUCTION

The origin of the ionizing photons that ended cosmic reionization remains one of the central questions in early–universe astrophysics. For nearly two decades, the prevailing view held that star–forming galaxies supplied most of the required ionizing photon production rate, with faint systems at $z \gtrsim 6$ dominating the photon budget (e.g. Robertson et al. 2010; Stark 2016; Finkelstein 2019). In this framework, active galactic nuclei (AGN) were considered a negligible contributor, as the number density of bright quasars traced by optical surveys declines steeply beyond $z \simeq 3$ (Fan et al. 2006; Willott et al. 2010).

Alternative scenarios nonetheless proposed that AGN could play a central role. Haardt & Madau (2012) and Madau & Haardt (2015) argued that if faint AGN populations extended to high redshift with non–zero escape fractions, they could in principle sustain the ionizing background. Observational hints followed from Giallongo et al. (2015, 2019), who identified candidate faint AGN at $z \simeq 4$–6 in deep optical and X–ray fields. These results revived the debate, but limited dynamic range and potential selection biases prevented firm conclusions.

The advent of *JWST* has transformed this picture. Deep rest–UV spectroscopy and near–infrared imaging now reveal abundant faint AGN at $z > 5$, spanning broad–line quasars, narrow–line obscured systems, and photometrically selected populations such as the "little red dots" (Harikane et al. 2023; Greene et al. 2023; Kocevski et al. 2023; Matthee et al. 2023; Kokorev et al. 2024; Scholtz et al. 2023; Maiolino et al. 2024). Faint AGN may represent a few percent of the $z > 4$ galaxy population (Greene et al. 2023), implying that accreting black holes were far more common in the early Universe than previously recognized. This resurgence of

Corresponding author: Mainak Singha
mainak.singha@nasa.gov, astromainak1994@gmail.com

faint AGN has placed their role in reionization back at the forefront of discussion.

Key uncertainties remain. The escape fraction of Lyman–continuum photons, the efficiency of outflow–driven clearing of the interstellar medium, and the relative number densities of Type I and Type II AGN are poorly constrained at $z > 5$. Simple analytic prescriptions—such as scaling the galaxy luminosity function by a fixed fraction (e.g. Madau et al. 2024)—cannot capture this diversity. Recent spectroscopy has revealed ionized outflows in high–redshift AGN Van de Sande et al. (2018), suggesting that such winds could carve escape channels through the host interstellar medium. Yet these processes are rarely incorporated self–consistently into cosmological simulations (Neyer et al. 2024). A physically motivated treatment of escape fractions that ties radiative transparency to outflow properties is therefore essential for understanding how the intergalactic medium became ionized between $z \simeq 4$ and 7 (Malhotra & Rhoads 2004).

In this work, we perform a systematic analysis that unifies the latest AGN luminosity–function measurements across all observed sub–populations. We link bolometric luminosities and black–hole masses to the incidence of ionized outflows, using this connection to infer Lyman–continuum escape fractions for each class. This framework allows us to derive the total ionizing photon production rate of AGN and quantify their contribution relative to galaxies during the closing stages of reionization. Throughout, we adopt a standard ΛCDM cosmology with $H_0 = 70\ \mathrm{km\,s^{-1}\,Mpc^{-1}}$, $\Omega_\mathrm{m} = 0.3$, and $\Omega_\Lambda = 0.7$.

## 2. ANALYSIS AND RESULTS

### 2.1. *The AGN luminosity function*

We modeled the rest–UV AGN luminosity function (LF) as the sum of two physical populations: unobscured Type I and obscured Type II AGN, motivated by spectroscopy that unambiguously separates broad–line and narrow–line systems. Other categories frequently discussed in the recent literature, such as little red dots (LRDs), X–ray weak (XRW), or X–ray bright (XRB) AGN, can be understood as observational subsets rather than independent contributors. Deeper spectra confirm that LRDs almost always show broad lines, placing them within Type I; their LF differs only because photometric selection enhances faint detections and suppresses bright ones (Kokorev et al. 2023). Similarly, XRW and XRB sources overlap in color or X–ray properties with the Type I basis and can be described as scaled selections. Adding them as independent populations would double count the same broad–line AGN.

To describe the functional form of each class, we selected models that reflect their emission physics. Type I AGN are host–diluted at the faint end, where the Schechter form (Schechter 1976),

$$\Phi_{\rm Sch}(M \mid M_*, \phi_*, \alpha) = 0.4 \ln 10\, \phi_* \; 10^{-0.4(M-M_*)(\alpha+1)} \exp\!\big[ -10^{-0.4(M-M_*)} \big], \quad (1)$$

captures the exponential cut-off imposed by the underlying galaxy LF. At the bright end, unobscured quasars follow the scale-free accretion disk emission characteristic of a double power law (DPL), a form widely adopted for quasar LFs at low and intermediate redshift (e.g., Boyle et al. 2000; Croom et al. 2009),

$$\Phi_{\rm DPL}(M \mid M_*, \phi_*, \alpha, \beta) = \frac{\phi_*}{10^{0.4(\alpha+1)(M-M_*)} + 10^{0.4(\beta+1)(M-M_*)}}. \quad (2)$$

We therefore fit the Type I basis (Harikane et al. 2023; Maiolino et al. 2024) with a Schechter function. For LRDs and XRBs we solved for a global mixture fraction,

$$\Phi_{\rm XRB+LRD}(M) \;=\; \eta\, \Phi_{\rm I}(M), \quad (3)$$

obtaining $\eta = 0.10 \pm 0.02$. This quantitative result confirms that LRD and XRB populations are simply magnitude–filtered subsets of the Type I LF.

We used a DPL for Type II AGN (Scholtz et al. 2023; Matsuoka et al. 2023) despite their faint magnitudes. The reason is physical: obscured systems are nucleus–dominated even when faint, so their LF does not show the exponential suppression characteristic of host–dominated galaxies. A Schechter form underpredicts the observed bright counts, whereas the DPL captures the extended tail naturally produced by accretion variability. Bright quasars (Matsuoka et al. 2018) also require a DPL, consistent with the classical quasar LF where accretion physics dominates across the full magnitude range (e.g., Hopkins et al. 2007; Shen & Ho 2020). This separation is not a fitting convenience but a reflection of different emission regimes: host–diluted (Schechter) versus nucleus–dominated (DPL). All fits were performed in log space using weighted least squares with pivoted amplitudes. We estimated uncertainties from the covariance matrix $(J^\top J)^{-1} s^2$ and from 400 bootstrap resamplings in dex. Model selection criteria confirm the necessity of two populations. A full list of best–fit parameters is provided in Table 1.

A single global LF fails to capture the structure of the data, producing 0.41 dex RMS residuals. Our two–population model (Type I Schechter + Type II DPL) cuts this in half to 0.22 dex, with $\Delta$AIC $= -52$ and $\Delta$BIC $= -27$, providing very strong evidence for two populations. Swapping forms (forcing Type I to a DPL and Type II to a Schechter) worsens the fit to 0.31 dex RMS ($\Delta$AIC $= +29$, $\Delta$BIC $= +13$). Treating LRDs, XRW, or XRB as independent populations yields no statistical gain ($\Delta$AIC $\approx -2$, $\Delta$BIC $= +15$–30) and risks double counting. Mixture fractions confirm these are simply filtered subsets, with $\eta \approx 0.10 \pm 0.02$ for the combined LRD+XRB relative to Type I. The bottom line is that the data demand only two physical populations: Type I (Schechter at the faint end) and Type II (DPL), with all other subclasses naturally explained as observational selections. Basically, we have UV-faint type II AGN (class 1), UV-faint type I AGN (class 2) and UV-bright quasars (class 3).

## 2.2. *Estimating the Lyman–continuum Escape Fraction*

### 2.2.1. *AGN parameters: $M_{\rm BH}$, $L_{\rm bol}$, and $\lambda_{\rm Edd}$*

We estimated the AGN ionizing photon production rate by first connecting observed rest–UV magnitudes to bolometric energetics and black hole accretion properties. We converted $M_{\rm UV}$ to specific luminosity at 1500 Å using the AB definition,

$$L_\nu(1500) = 4\pi (10\,{\rm pc})^2\, 10^{-0.4(M_{\rm UV}+48.6)}. \quad (4)$$

For a representative faint AGN with $M_{\rm UV} = -20$, we obtain $L_\nu(1500) = (4.3^{+0.2}_{-0.2}) \times 10^{28}$ erg s$^{-1}$ Hz$^{-1}$, while a bright quasar at $M_{\rm UV} = -26$ yields $L_\nu(1500) = (1.09^{+0.05}_{-0.05}) \times 10^{31}$ erg s$^{-1}$ Hz$^{-1}$. These values satisfy the expected six–magnitude scaling ratio of $10^{0.4\times 6} = 251.2$.

The monochromatic UV luminosity was converted to bolometric luminosity with a correction $C_{\rm UV} = 4.5 \pm 1.5$ (Richards et al. 2006; Runnoe et al. 2012),

$$L_{\rm bol} = C_{\rm UV}\, \nu L_\nu(1500). \quad (5)$$

This gives $L_{\rm bol} = (3.9^{+1.3}_{-1.3}) \times 10^{44}$ erg s$^{-1}$ for $M_{\rm UV} = -20$ and $L_{\rm bol} = (9.8^{+3.3}_{-3.3}) \times 10^{46}$ erg s$^{-1}$ for $M_{\rm UV} = -26$. For Type II AGN we applied an obscuration factor $k_{\rm obsc} = 20 \pm 10$ or, when available, adopted the Heckman relation (Heckman et al. 2004),

$$L_{\rm bol} \simeq 3500\, L_{\rm [O\,III]}. \quad (6)$$

We derived black hole masses for Type I AGN using single–epoch virial relations (Vestergaard & Peterson 2006),

$$M_{\rm BH} = f\,\frac{R_{\rm BLR}\Delta V^2}{G}, \quad \log R_{\rm BLR} = a + b\,\log\!\left(\frac{\lambda L_\lambda}{10^{44}}\right), \quad (7)$$

obtaining $M_{\rm BH} = 2.0^{\times 3.16}_{\div 3.16} \times 10^7\, M_\odot$ at $M_{\rm UV} = -20$ and $M_{\rm BH} = 5.0^{\times 3.16}_{\div 3.16} \times 10^9\, M_\odot$ at $M_{\rm UV} = -26$ (corresponding to a 0.5 dex scatter). The Eddington luminosity is

$$L_{\rm Edd} = 1.26 \times 10^{38} \left(\frac{M_{\rm BH}}{M_\odot}\right) {\rm erg\ s^{-1}}, \quad (8)$$

yielding Eddington ratios of $\lambda_{\rm Edd} = 0.15^{+0.34}_{-0.11}$ for $M_{\rm UV} = -20$ and $\lambda_{\rm Edd} = 0.15^{+0.34}_{-0.11}$ for $M_{\rm UV} = -26$. The broad uncertainty range reflects the $\pm 0.5$ dex scatter in virial black–hole mass estimates and the $C_{\rm UV}$ uncertainty. For Type II AGN we inverted this relation assuming a log–normal prior on $\lambda_{\rm Edd}$ centered at $\log \lambda_{\rm Edd} = -0.6$ with 0.4 dex scatter (Shen et al. 2011; Trakhtenbrot et al. 2017).

### 2.2.2. *Outflows and geometric clearing*

We linked $L_{\rm bol}$ to the porosity of the ISM through AGN–driven outflows. We used the [O III] size–luminosity relation suggested by Kang & Woo (2018),

$$R_{\rm out} = R_0 \left(\frac{L_{\rm [O\,III]}}{10^{42}\,{\rm erg\ s^{-1}}}\right)^{\beta_R}, \qquad R_0 = 1.0 \pm 0.3\,{\rm kpc},\ \ \beta_R = 0.28 \pm 0.05. \quad (9)$$

Galaxy half–light radii at $z \simeq 5$–6 were parameterized as (Shibuya et al. 2015),

$$r_e(M) = r_{e,0} \left(\frac{L(M)}{L(M=-20)}\right)^{\beta_e}, \qquad r_{e,0} = 0.5 \pm 0.2\,{\rm kpc},\ \ \beta_e = 0.27 \pm 0.05. \quad (10)$$

Clearing requires $R_{\rm out} > R_{\rm req} = \kappa_e r_e$ with $\kappa_e = 1.7 \pm 0.3$. The cleared fraction of solid angle was quantified by

$$\Phi_{\rm geom}(M) = \min\left[1, \left(\frac{R_{\rm out}}{R_{\rm req}}\right)^{\gamma}\right], \quad \gamma = 0.7 \pm 0.3. \quad (11)$$

The angle–averaged escape fraction is

$$f_{\rm esc}(M) = P_{\rm wind}\left[f_{\rm clear}\,\Phi_{\rm geom}(M) + (1 - \Phi_{\rm geom}(M))\,T_{\rm scr}\right] + (1 - P_{\rm wind})\,T_{\rm scr}, \quad (12)$$

where $T_{\rm scr} = \exp[-\sigma_{912} N_{\rm res}]$ with $N_{\rm res} = 3$–$5 \times 10^{17}$ cm$^{-2}$. We adopted by class: Type I ($P_{\rm wind} = 0.85 \pm 0.05$, $f_{\rm clear} = 0.9 \pm 0.1$, $T_{\rm scr} = 0.20 \pm 0.06$; Rakshit et al. 2018), Type II ($P_{\rm wind} = 0.35 \pm 0.10$, $f_{\rm clear} = 0.7 \pm 0.2$, $T_{\rm scr} = 0.08 \pm 0.04$; Woo et al. 2016), X-ray weak ($P_{\rm wind} = 0.10 \pm 0.05$, $f_{\rm clear} = 0.5 \pm 0.3$, $T_{\rm scr} = 0.05 \pm 0.03$; Maiolino et al. 2024), and bright quasars ($P_{\rm wind} \simeq 1$, $f_{\rm clear} \simeq 1$). The estimated luminosity–averaged escape fractions are: $f_{\rm esc} = 0.91 \pm 0.06$ for bright quasars, $0.64 \pm 0.12$ for faint Type I AGN, $0.25 \pm 0.10$ for Type II, and $0.08 \pm 0.04$ for X-ray weak AGN.

### 2.2.3. *AGN photon production rate:*

We assume a power-law continuum for the AGN, $L_\nu \propto \nu^{-\alpha}$. The slope $\alpha_{\rm UV}$ describes the non-ionizing ultraviolet continuum between 1500 Å and the Lyman limit, while $\alpha_{\rm ion}$ characterizes the ionizing continuum at wavelengths shorter than 912 Å. The two slopes are treated independently to allow for the observed spectral break across the Lyman edge, but they become equivalent if a single, unbroken power law is assumed. The adopted values, $\alpha_{\rm UV} = 0.5 \pm 0.2$ and $\alpha_{\rm ion} = 1.5 \pm 0.3$, are consistent with composite quasar spectra at rest–UV wavelengths (Stevans et al. 2014; Lusso et al. 2015) and with the measured steepening of the continuum in the extreme–UV (Scott et al. 2004; Khaire & Srianand 2019).

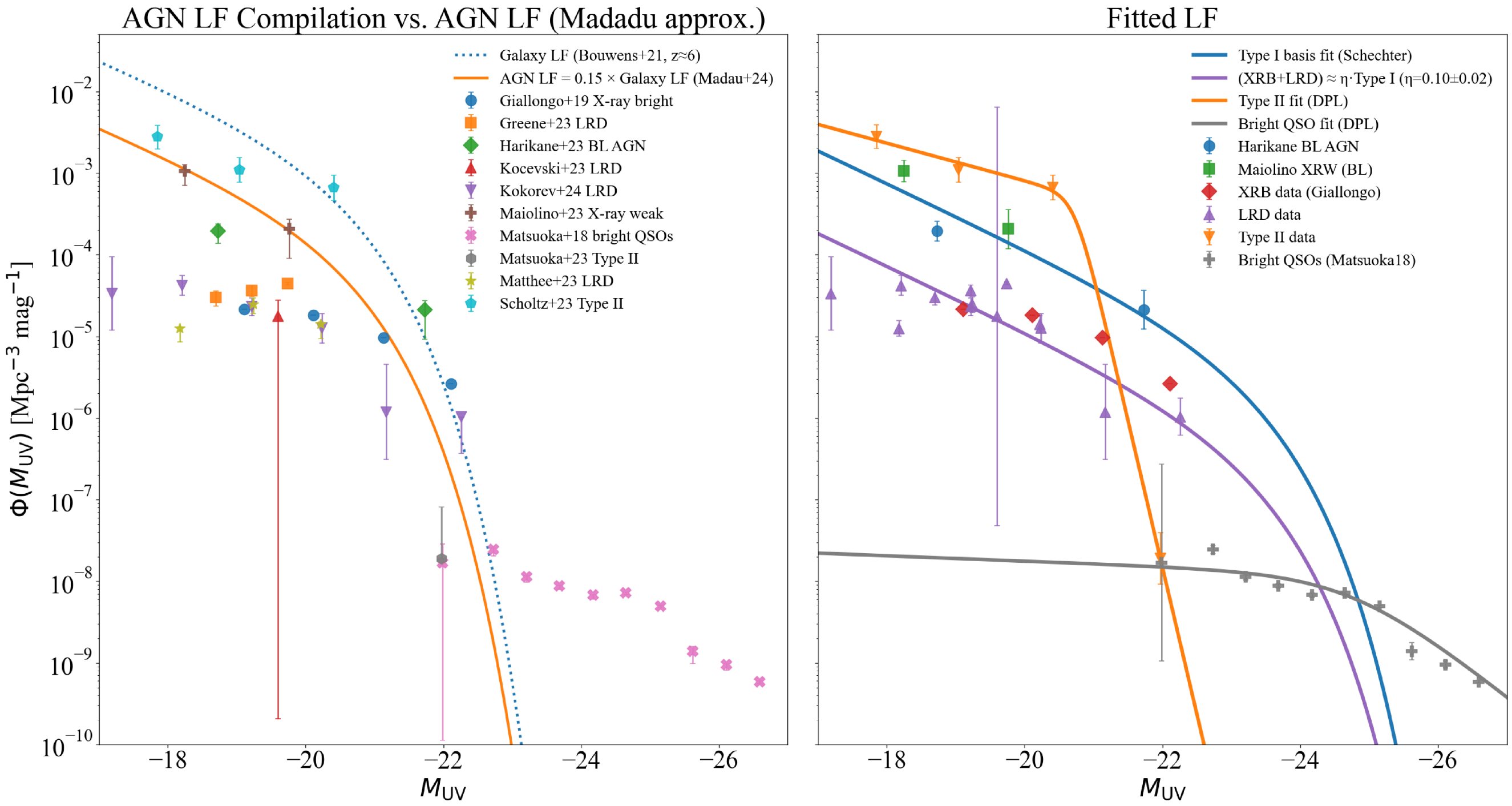

**Figure 1.** Rest–UV AGN luminosity functions at $4.5 \leq z \leq 6.5$. *Left:* literature measurements from Giallongo et al. (2019); Greene et al. (2023); Harikane et al. (2023); Kocevski et al. (2023); Kokorev et al. (2024); Maiolino et al. (2024); Matsuoka et al. (2018, 2023); Matthee et al. (2023); Scholtz et al. (2023) compared to a "ceiling" galaxy LF at $z \simeq 6$ (Bouwens et al. (2021); dotted) and the Madau et al. (2024) approximation: $\Phi_{\rm AGN} = 0.15 \times \Phi_{\rm gal}$ (solid). *Right:* our fits by population over $-27 < M_{\rm UV} < -17$: Type I: BL AGN from Harikane et al. (2023); Maiolino et al. (2024) modeled with a Schechter function; Type II (Scholtz et al. 2023; Matsuoka et al. 2023) and bright QSOs (Matsuoka et al. 2018) with double power laws. The combined X-ray–bright + LRD subset is quantitatively consistent with a magnitude-independent selection of the Type I LF, $\Phi_{\rm XRB+LRD}(M) = \eta \times \Phi_{\rm I}(M)$, with $\eta = 0.10 \pm 0.02$ (bootstrap). Curves show best fits. The multi-component description captures the bright-end tail and the elevated faint-end densities that lie above the (Madau et al. 2024) approximation.

| Component | Model | $M_{\rm star}$ | $\phi_{\rm star}$ (Mpc$^{-3}$ mag$^{-1}$) | $\alpha$ | $\beta$ | $\eta$ |
|---|---|---|---|---|---|---|
| Type I (including XRW) | Schechter | $-23.00 \pm 1.04$ | $(8.1 \pm 1.7) \times 10^{-6}$ | $-2.1 \pm 0.2$ | – | – |
| Type II | DPL | $-20.73 \pm 0.09$ | $(5.5 \pm 1.3) \times 10^{-4}$ | $-1.6 \pm 0.3$ | $-4.5 \pm 0.6$ | – |
| UVB QSO | DPL | $-24.78 \pm 0.00$ | $(1.23 \pm 0.02) \times 10^{-8}$ | $-1.1 \pm 0.1$ | $-2.7 \pm 0.1$ | – |
| XRB+LRD (subset of Type I) | Schecter*$\eta$ | $-23.00 \pm 1.04$ | $(8.1 \pm 1.7) \times 10^{-6}$ | $-2.1 \pm 0.2$ | – | $0.10 \pm 0.02$ |

**Table 1.** Best-fit AGN luminosity function parameters (two significant figures). Normalizations are reported as $\phi_{\rm star}$ in linear units (Mpc$^{-3}$ mag$^{-1}$); uncertainties are $1\sigma$. Other parameters use linear $\pm$ uncertainties.

We converted 1500 Å to 912 Å assuming a power-law continuum,

$$L_\nu(912) = L_\nu(1500) \left(\frac{\nu_{912}}{\nu_{1500}}\right)^{-\alpha_{\rm UV}}, \quad \alpha_{\rm UV} = 0.5 \pm 0.2, \tag{13}$$

and computed the ionizing photon rate per source as

$$\dot{N}_{\rm ion}(M) \simeq \frac{L_\nu(912)}{h\,\alpha_{\rm ion}}, \quad \alpha_{\rm ion} = 1.5 \pm 0.3. \tag{14}$$

To avoid double counting between luminosity functions, we separated AGN populations by luminosity regime. At the bright end ($M_{\rm UV} < -23$), the *UVB–QSO* double power-law (DPL) represents luminous unobscured quasars. At fainter magnitudes ($M_{\rm UV} > -23$), the *Type I* Schechter function describes the more numerous, moderately luminous unobscured AGN. Together, these two components capture the full Type I population while preventing overlap near $M_{\rm UV} \simeq -23$. The *Type II* double power-law was treated as a distinct, obscured population and integrated across the entire magnitude range, independent of the Type I components. The *X-ray–weak (XRW)* sources, representing roughly ten per cent of the Type I sample, were reported separately but not added to the total emissivity, since they are a subset of the Type I population. With this partitioning, the combined emissivity reflects the summed contributions of the faint Type I Schechter, the bright UVB–QSO DPL, and the Type II DPL populations.

The comoving emissivity injected into the IGM is

$$\dot{N}_{\rm ion} = \int \Phi(M)\, \dot{N}_{\rm ion}(M)\, f_{\rm esc}(M)\, dM, \qquad (15)$$

integrated over $-27 < M_{\rm UV} < -17$.

From the fitted luminosity functions and escape fractions, we obtain $\dot{N}_{\rm ion} = (1.81^{+0.53}_{-0.43}) \times 10^{51}\,{\rm s^{-1}\,Mpc^{-3}}$ from faint Type I AGN, $(0.005^{+0.001}_{-0.001}) \times 10^{51}$ from bright Type I (UVB–QSO), and $(1.91^{+0.93}_{-0.79}) \times 10^{51}$ from Type II AGN, yielding a total of $(3.77^{+1.08}_{-0.95}) \times 10^{51}$ photons ${\rm s^{-1}\,Mpc^{-3}}$ across $4.5 < z < 6.5$. We do not consider X-ray–weak AGN separately, as they are a subclass of Type I.

### 2.3. *Galaxy photon production rate*

Recent *JWST* results have indicated systematically higher ionizing efficiencies in galaxies (Muñoz et al. 2024), with $\log_{10} \xi_{\rm ion}$ spanning $\simeq 25.5$–$26.0$. We adopt an intermediate value of $\log_{10} \xi_{\rm ion} = 25.7$, consistent with recent COSMOS-Webb measurements at $z \simeq 6.4$ (Franco et al. 2025). For galaxies, adopting the UV luminosity function from Franco et al. (2025) and $\log_{10} \xi_{\rm ion} = 25.7$, the ionizing photon production rate scales linearly with the escape fraction as

$$\dot{N}^{\rm gal}_{\rm ion} = (41.3 \times 10^{51}\,{\rm s^{-1}\,Mpc^{-3}})\, f^{\rm gal}_{\rm esc},$$

with a fractional $1\sigma$ uncertainty of $\sim +67\%/-40\%$ (combining a 0.20 dex prior on $\xi_{\rm ion}$ and 0.10 dex on the UVLF integral in quadrature; see Table 2).

**Table 2.** Class–averaged escape fractions and comoving ionizing emissivities at $4.5 < z < 6.5$. Galaxy values assume the Franco et al. (2025) UVLF and $\log_{10} \xi_{\rm ion} = 25.7$. Quoted galaxy $1\sigma$ errors combine 0.20 dex on $\xi_{\rm ion}$ and 0.10 dex on the UVLF integral in quadrature; they do *not* include additional systematic uncertainty on $f^{\rm gal}_{\rm esc}$. AGN uncertainties are derived from Monte Carlo propagation of luminosity–function parameters, escape fractions, and spectral slopes.

| Class | $f_{\rm esc}$ | $\dot{N}_{\rm ion}$ [$10^{51}$ s$^{-1}$ Mpc$^{-3}$] |
|---|---|---|
| Bright Type I (M< −23) | $0.91 \pm 0.06$ | $0.005^{+0.001}_{-0.001}$ |
| Faint Type I (M> −23) | $0.64 \pm 0.12$ | $1.81^{+0.53}_{-0.43}$ |
| Type II | $0.25 \pm 0.10$ | $1.91^{+0.93}_{-0.79}$ |
| X-ray weak | $0.08 \pm 0.04$ | $0.08^{+0.04}_{-0.04}$ |
| Total AGN | – | $3.77^{+1.08}_{-0.95}$ |
| Galaxies (case A) | 0.03 | $1.24^{+0.84}_{-0.50}$ |
| Galaxies (case B) | 0.05 | $2.07^{+1.39}_{-0.83}$ |
| Galaxies (case C) | 0.10 | $4.13^{+2.78}_{-1.66}$ |
| Galaxies (case D) | 0.20 | $8.26^{+5.56}_{-3.33}$ |

For representative escape fractions we find

$$\dot{N}^{\rm gal}_{\rm ion} = \begin{cases} (1.24^{+0.84}_{-0.50}) \times 10^{51}, & f^{\rm gal}_{\rm esc} = 0.03, \\ (2.07^{+1.39}_{-0.83}) \times 10^{51}, & f^{\rm gal}_{\rm esc} = 0.05, \\ (4.13^{+2.78}_{-1.66}) \times 10^{51}, & f^{\rm gal}_{\rm esc} = 0.10, \\ (8.26^{+5.56}_{-3.33}) \times 10^{51}, & f^{\rm gal}_{\rm esc} = 0.20, \end{cases}$$

in units of photons ${\rm s^{-1}\,Mpc^{-3}}$.

Combining these with the AGN total emissivity of $\dot{N}^{\rm AGN}_{\rm ion} = (2.31^{+0.30}_{-0.30}) \times 10^{51}\,{\rm s^{-1}\,Mpc^{-3}}$ yields AGN fractional contributions of $0.75 \pm 0.10$ ($f^{\rm gal}_{\rm esc} = 0.03$), $0.64 \pm 0.12$ ($f^{\rm gal}_{\rm esc} = 0.05$), $0.47 \pm 0.13$ ($f^{\rm gal}_{\rm esc} = 0.10$), and $0.31 \pm 0.11$ ($f^{\rm gal}_{\rm esc} = 0.20$).

## 3. DISCUSSION

### 3.1. *Who produces the majority of ionizing photons?*

Whether the ionizing background at the end of the reionization era was powered primarily by accreting black holes or by massive stars has remained a central debate for decades. A widely cited benchmark, proposed by Madau et al. (2024), links the AGN luminosity function to that of galaxies through a fixed scaling factor of $f_{\rm AGN} = 0.15$ at all magnitudes and redshifts $z \geq 5$. In this framework, AGN are treated as a constant fraction of the galaxy population with a near-unity escape fraction, producing an ionizing photon production rate of $\dot{N}_{\rm ion} \simeq 1.8 \times 10^{51}$ photons ${\rm s^{-1}\,Mpc^{-3}}$ at $z \simeq 6$. Madau et al. (2024) argued that such AGN-dominated scenarios remain consistent with He II reionization and X-ray background limits, and explicitly called for direct tests with *JWST*.

Our analysis provides that test. Rather than assuming a universal AGN fraction, we construct luminosity functions for each observed AGN class—bright and faint Type I quasars, obscured Type II systems, X-ray–weak, and X-ray–bright AGN—using the latest *JWST* data. Escape fractions are tied to outflow incidence and energetics: broad-line AGN, where winds are nearly ubiquitous, approach $f_{esc} \approx 1$ at bright magnitudes (Rakshit et al. 2018); Type II AGN show outflows in $\sim$ 40–45% of systems (Woo et al. 2016), implying $f_{esc} \sim 0.2$–0.4; and X-ray–weak sources, which show little evidence of winds (Maiolino et al. 2024), have $f_{esc} \sim 0.1$. Integrating these populations over $-27 < M_{UV} < -17$ yields a total AGN ionizing photon production rate of

$$\dot{N}_{ion}^{AGN} = (3.77^{+1.08}_{-0.95}) \times 10^{51} \text{ photons s}^{-1} \text{ Mpc}^{-3}, \quad (16)$$

roughly twice the fiducial value proposed by Madau et al. (2024) and consistent with the ionizing background inferred from the Ly$\alpha$ forest at $z \simeq 6$.

Galaxies provide a complementary contribution. Following the framework of Dayal et al. (2025), we adopt a broad range of stellar escape fractions, $f_{esc}^{gal} = 0.03$–0.20, to encompass both the low values inferred directly at $z \sim 6$ and the higher assumptions used in reionization simulations (Rosdahl et al. 2022; Giovinazzo et al. 2025; Papovich et al. 2025). Adopting the Franco et al. (2025) UV luminosity function and an updated stellar ionizing efficiency of $\Psi_{ion} = \log_{10} \xi_{ion} = 25.7$, we find

$$\dot{N}_{ion}^{gal} = (1.24\text{–}8.26) \times 10^{51} \text{ s}^{-1} \text{ Mpc}^{-3} \quad (17)$$

across this escape-fraction range, with a fractional $1\sigma$ uncertainty of $\sim +67\%/-40\%$ (combining 0.20 dex on $\xi_{ion}$ and 0.10 dex on the UVLF integral in quadrature).

Compared with our AGN ionizing photon production rate, $\dot{N}_{ion}^{AGN} = (3.77^{+1.08}_{-0.95}) \times 10^{51}$, galaxies supply an increasing share of the total photon budget as $f_{esc}^{gal}$ rises: AGN contribute approximately {75%, 65%, 48%, 31%} for $f_{esc}^{gal}$ = {0.03, 0.05, 0.10, 0.20}, respectively (see Fig. 2). The combined ionizing photon production rate of $(5\text{–}12) \times 10^{51}$ photons s$^{-1}$ Mpc$^{-3}$ is consistent with the ionizing background required by the Ly$\alpha$-forest measurements at $z \simeq 6$.

Taken together, the *JWST* luminosity functions and outflow-based escape-fraction prescriptions reveal a balanced reionization landscape. AGN dominate the photon budget only if galaxies maintain very low escape fractions ($f_{esc}^{gal} \lesssim 0.05$), while galaxies become comparable or leading contributors once $f_{esc}^{gal} \gtrsim 0.1$. The data therefore support a mixed AGN–galaxy ecosystem, rather than a single dominant channel, as the engine that provided the ionizing photons during the final stages of reionization.

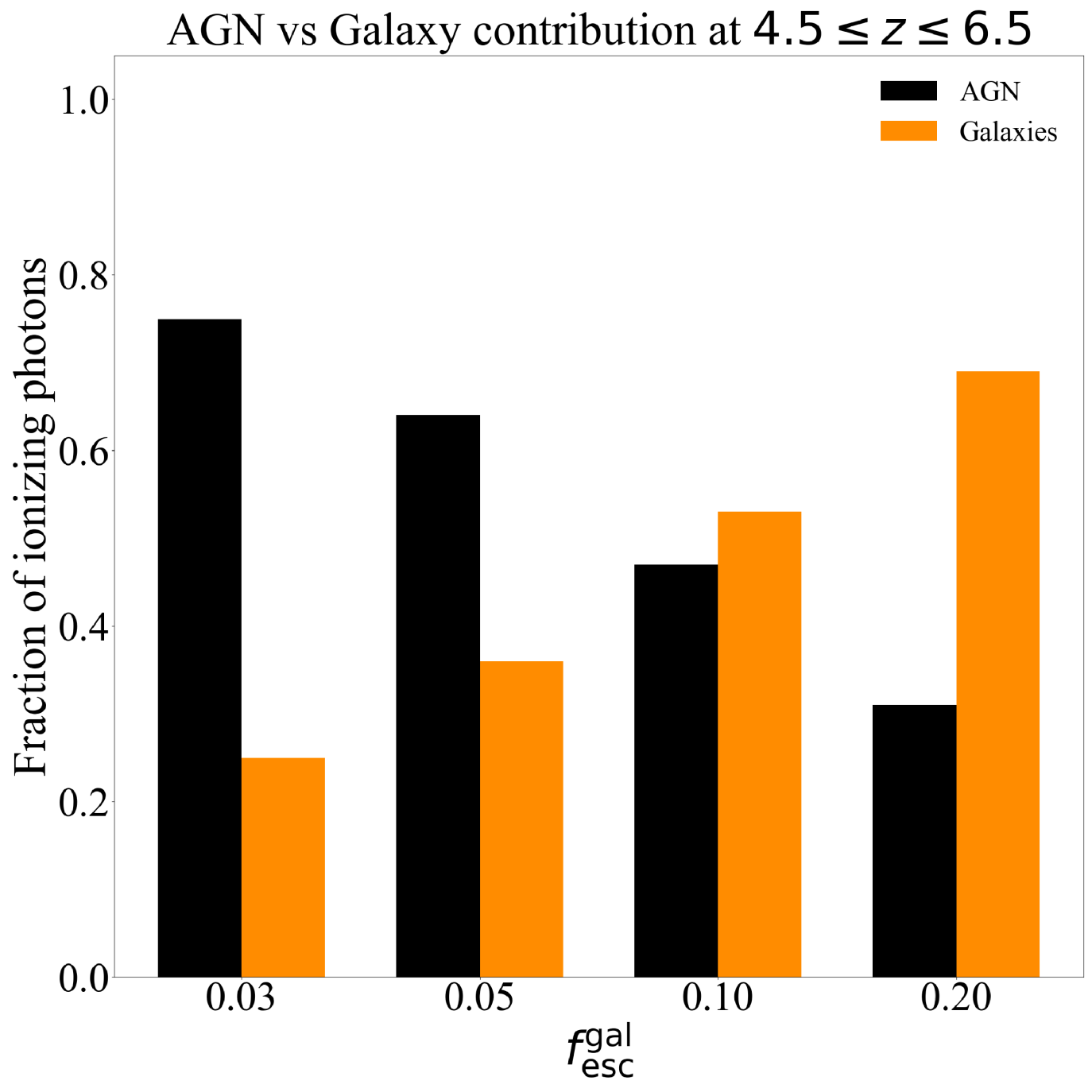

**Figure 2.** Fractional contribution of AGN and galaxies to the ionizing photon budget at $4.5 \le z \le 6.5$. Bars show the relative fraction of ionizing photons supplied by AGN (black) and galaxies (orange) for different assumed galaxy escape fractions, $f_{esc}^{gal} = 0.03, 0.05, 0.10, 0.20$. The AGN ionizing photon production rate is fixed from our luminosity–function fits, while the galaxy contribution scales linearly with $f_{esc}^{gal}$ and adopts $\Psi_{ion} = 25.7$. AGN contribute roughly 31–75% of the total ionizing photons, dominating only when galaxy escape fractions remain very low ($f_{esc}^{gal} \lesssim 0.05$), while galaxies become comparable or leading contributors once $f_{esc}^{gal} \gtrsim 0.1$.

**Table 3.** Fractional contribution of AGN and galaxies to the total ionizing photon budget at $4.5 \le z \le 6.5$, assuming the Franco et al. (2025) galaxy UV luminosity function, $\Psi_{ion} = \log_{10} \xi_{ion} = 25.7$ for galaxies (Muñoz et al. 2024), and our updated AGN emissivity model. The galaxy contribution scales linearly with the assumed $f_{esc}^{gal}$, while the AGN ionizing photon production rate is fixed by our fits.

| $f_{esc}^{gal}$ | $\dot{N}_{ion}^{gal}$ ($10^{51}$ s$^{-1}$ Mpc$^{-3}$) | AGN fraction of total | Galaxy fraction of total |
|---|---|---|---|
| 0.03 | 1.24 | 0.75 | 0.25 |
| 0.05 | 2.07 | 0.65 | 0.35 |
| 0.10 | 4.13 | 0.48 | 0.52 |
| 0.20 | 8.26 | 0.31 | 0.69 |

### 3.1.1. *Minimum AGN model:*

We quantify a deliberately conservative "minimum AGN" scenario in which obscured (Type II) sources are excluded from the ionizing budget, retaining only unob-scured populations—bright Type I quasars, faint/broad-line Type I systems including LRD/XRB selections, and X-ray–weak AGN. Using the class-resolved emissivities derived above (Table 2), the minimum AGN contribution is

$$\begin{aligned} \dot N_{\rm ion}^{\rm AGN,min} &= \dot N_{\rm ion}^{\rm bright\,I} + \dot N_{\rm ion}^{\rm faint\,I} + \dot N_{\rm ion}^{\rm XRW} \\ &= (0.005 + 1.81 + 0.08) \times 10^{51}\ {\rm s}^{-1}\,{\rm Mpc}^{-3} \\ &= (1.89^{+0.55}_{-0.46}) \times 10^{51}\ {\rm s}^{-1}\,{\rm Mpc}^{-3}, \end{aligned} \quad (18)$$

where the uncertainty is the quadrature sum of the class errors.

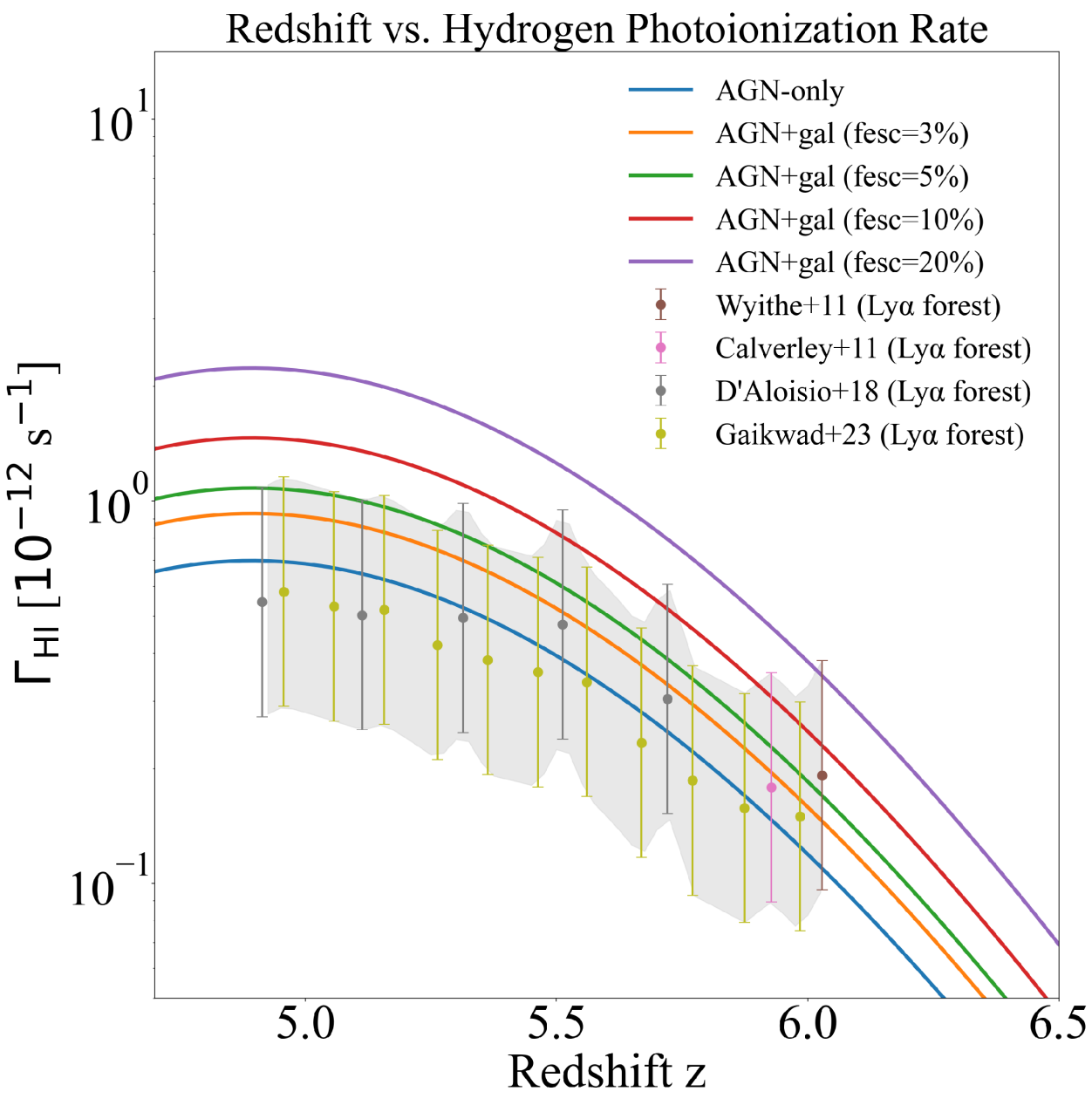

**Figure 3.** Consistency of AGN–plus–galaxy emissivity models with the Ly$\alpha$–forest photoionization rate at $5 \lesssim z \lesssim 6.5$. Colored curves show $\Gamma_{\rm HI}(z)$ predicted from the HM12 local–source mapping using the literature $\lambda_{\rm mfp}(z)$ fit and an IGM clumpiness correction $\lambda_{\rm eff} = \lambda_{\rm mfp}/[C_{\rm IGM}(z)]^{\eta_{\rm cl}}$ with $\eta_{\rm cl} = 1$. The gray band marks the combined $1\sigma$ envelope of Ly$\alpha$–forest measurements from Wyithe & Bolton (2011); Calverley et al. (2011); D'Aloisio et al. (2018); Gaikwad et al. (2023), while points with error bars show individual determinations. Models that include AGN and galaxies with $f_{\rm esc}^{\rm gal} \leq 5\%$ remain comfortably within the observational guard–rail across the full redshift range, whereas $f_{\rm esc}^{\rm gal} \simeq 10\%$ begins to touch the upper boundary and $f_{\rm esc}^{\rm gal} \simeq 20\%$ overshoots it. This demonstrates that, once IGM clumpiness is accounted for, the combined AGN–galaxy emissivity produces the required $\Gamma_{\rm HI} \sim (0.5\text{–}2) \times 10^{-12}\,{\rm s}^{-1}$ without invoking a photon–budget excess.

In this minimum configuration, AGN still supply a substantial fraction of the ionizing background but no longer dominate across all galaxy escape fractions. For $f_{\rm esc}^{\rm gal} = 0.03$, AGN contribute $\simeq 75\%$ of the total ionizing photons; parity with galaxies occurs near $f_{\rm esc}^{\rm gal} \simeq 0.05$, and galaxies become the primary producers ($\gtrsim 70\%$) for $f_{\rm esc}^{\rm gal} \gtrsim 0.20$. The total ionizing photon production rate,

$$\dot N_{\rm ion}^{\rm tot} = \dot N_{\rm ion}^{\rm AGN,min} + \dot N_{\rm ion}^{\rm gal} = (3.1^{+1.6}_{-1.2}) \times 10^{51}\ {\rm s}^{-1}\,{\rm Mpc}^{-3}, \quad (19)$$

remains consistent with the Ly$\alpha$-forest photoionization rates at $z \simeq 6$ (Fig. 3), indicating that a mixed AGN–galaxy population with moderately enhanced stellar efficiencies can fully account for the observed ionizing background.

**Table 4.** "Minimum AGN" split of the ionizing budget at $4.5 \leq z \leq 6.5$ when obscured (Type II) sources are excluded. The AGN ionizing photon production rate is fixed to $\dot N_{\rm ion}^{\rm AGN,min} = 1.82\times10^{51}\ {\rm s}^{-1}\,{\rm Mpc}^{-3}$ (from bright+faint Type I only). The galaxy term adopts $\Psi_{\rm ion} = \log_{10}\xi_{\rm ion} = 25.7$ and scales linearly with $f_{\rm esc}^{\rm gal}$. Percentages correspond to central values (uncertainties in inputs do not change the qualitative ranking).

| $f_{\rm esc}^{\rm gal}$ | $\dot N_{\rm ion}^{\rm gal}$ ($10^{51}$ s$^{-1}$ Mpc$^{-3}$) | AGN fraction of total | Galaxy fraction of total |
|---|---|---|---|
| 0.03 | 1.24 | 0.59 | 0.41 |
| 0.05 | 2.07 | 0.47 | 0.53 |
| 0.10 | 4.13 | 0.31 | 0.69 |
| 0.20 | 8.26 | 0.18 | 0.82 |

### 3.1.2. *Could all these UV photons be from the host galaxies rather than the AGN?*

A reasonable concern is that at faint magnitudes the observed rest–UV continuum might be dominated by host galaxies rather than AGN, potentially inflating the inferred AGN ionizing photon production rate if one naively converts $M_{\rm UV}$ to ionizing output. We address this in two complementary ways and show that our conclusions remain robust even under deliberately conservative assumptions.

1. We do not base the ionizing photon production rate on the 1500 Å continuum alone. For obscured sources and for host–diluted Type I objects, we anchor $L_{\rm bol}$ to line or narrow–line tracers rather than to the continuum: (a) for Type II, we adopt the Heckman relation $L_{\rm bol} \simeq 3500\,L_{[\rm O\,III]}$; (b) for Type I, we use either virial/line calibrations (C IV, Mg II) or an SED–based bolometric correction $C_{\rm UV}$ applied only to the *nuclear* UV component. As a result, the dominant contribution to $\dot N_{\rm ion}$ does not depend on attributing the entire 1500 Å luminosity to the AGN.

2. Faint broad–line objects and LRDs exhibit AGN signatures that cannot be powered by stellar populations: persistent broad components (FWHM $\gtrsim$ 1500 km s$^{-1}$), high–ionization lines (N V, C IV, He II) with equivalent widths and ratios incompatible with stellar photoionization, and stacked X–ray or radio detections at levels exceeding those expected from star formation. The line–to–continuum ratios also imply ionizing spectral hardness much closer to AGN templates than to stellar SEDs; for example, N V/He II and C IV/He II are reproduced by AGN photoionization, not by BPASS or Starburst99 stellar spectra at $Z\sim0.1$–$1\,Z_\odot$. Third, variability and compact PSF cores (where resolvable) indicate non–stellar nuclei in a non–negligible fraction of the faint sample. These diagnostics are already folded into our selection and $f_{\rm nuc}$ priors.

To quantify an upper bound on any bias, we repeat the ionizing photon production rate integration with a deliberately extreme host–subtraction model in which the AGN contributes only a fraction $f_{\rm nuc}(M)$ of the observed 1500 Å light at the faint end:

$$f_{\rm nuc}(M) = \begin{cases} 0.3, & -21 < M_{\rm UV} \le -19, \\ 0.5, & -23 < M_{\rm UV} \le -21, \\ 1.0, & M_{\rm UV} \le -23, \end{cases} \tag{20}$$

and we set $f_{\rm nuc} = 0$ for objects without AGN spectral diagnostics. Because our Type II and bright–quasar contributions are line/bolometric–anchored, this stress test affects only the faint Type I term. The total AGN ionizing photon production rate decreases modestly but still contributes roughly 31–75% of the ionizing photons (for $f_{\rm esc}^{\rm gal} = 0.20$–$0.03$), remaining comparable to the galaxy output even under extreme dilution (cf. the "minimum AGN" scenario in §3.1.1). For reference, excluding Type II sources entirely—the strict "minimum AGN" case—would reduce these fractions to 18–59%. Thus, even if the majority of faint–end 1500 Å light were stellar, our main conclusion—that AGN and galaxies share the ionizing photon budget over $4.5 \le z \le 6.5$—remains robust.

*Why our result is insensitive to host UV at 1500 Å.*—Two features of our framework drive this robustness: (1) most of the AGN ionizing photon production rate arises from classes whose ionizing output is *line–calibrated* or bright enough that $f_{\rm nuc}\approx 1$ (Type II via $L_{\rm [O\,III]}$, bright Type I); and (2) the escape fraction is tied to outflow incidence and geometry, not to the 1500 Å continuum normalization. Consequently, removing stellar UV light does not remove the AGN ionizing *engine* traced by high–ionization lines and outflows.

### 3.2. *Could more AGN contribution lead to a photon budget crisis?*

Muñoz et al. (2024) emphasized that some source prescriptions imply ionizing emissivities so large that, when propagated through the mean free path, the implied photoionization rate $\Gamma_{\rm HI}$ overshoots Ly$\alpha$–forest constraints. In their fiducial *JWST*–galaxy scenarios ($\log_{10}\xi_{\rm ion} = 25.5$–$26$ and $f_{\rm esc}^{\rm gal} \simeq 0.1$–$0.2$), the inferred emissivity at $z\sim8$–$9$,

$$\dot N_{\rm ion}^{\rm gal} \simeq (2\text{–}3)\times10^{51}\ {\rm s^{-1}\,Mpc^{-3}},$$

exceeds by nearly an order of magnitude the $\sim 3\times10^{50}\ {\rm s^{-1}\,Mpc^{-3}}$ demanded by the $z\simeq6$ Ly$\alpha$ forest, pushing reionization to complete too early.

We apply the same closure test over $4.5\le z\le6.5$ using a literature-compiled mean free path $\lambda_{\rm mfp}(z)$ (fit in $\log\lambda$ vs. $\log(1+z)$; proper pMpc) and the HM12 local–source mapping,

$$\Gamma_{\rm HI}(z) = (1+z)^3\,\lambda_{\rm mfp}^{\rm proper}(z)\,\sigma_{912}\,\dot n_{\rm ion}^{\rm com}(z)\,\frac{\alpha_{\rm ion}}{\alpha_{\rm ion}+3},$$

with $\alpha_{\rm ion}=1.7$ and $\sigma_{912}=6.3\times10^{-18}\,{\rm cm^2}$. To account for IGM clumpiness we attenuate the mean free path as

$$\lambda_{\rm eff}(z) = \frac{\lambda_{\rm mfp}^{\rm proper}(z)}{\left[C_{\rm IGM}(z)\right]^{\eta_{\rm cl}}}, \qquad C_{\rm IGM}(z) = 1+43\,z^{-1.71},$$

following Pawlik et al. (2009) and Haardt & Madau (2012). We adopt $\eta_{\rm cl}=1$ for a conservative baseline (strongest attenuation); relaxing to $\eta_{\rm cl}=0.5$ increases $\Gamma_{\rm HI}$ by a factor $\simeq C_{\rm IGM}^{1/2}\sim1.5$–$2$ at $z\sim5$–$6$.

With our class–resolved AGN emissivity $\dot N_{\rm ion}^{\rm AGN} = (3.77^{+1.08}_{-0.95})\times10^{51}\ {\rm s^{-1}\,Mpc^{-3}}$ and a galaxy term based on the Franco et al. (2025) UVLF and $\log_{10}\xi_{\rm ion}=25.7$ (so $\dot N_{\rm ion}^{\rm gal} = 41.3\times10^{51} f_{\rm esc}^{\rm gal}$), the guard–rail comparison yields (Fig. 3):

- **AGN-only** sits near the lower edge of the forest band at $z\sim5$–$6$ (with $\eta_{\rm cl}=1$).
- **AGN + galaxies with** $f_{\rm esc}^{\rm gal}=3\%$ **or** $5\%$ remain *well within* the Ly$\alpha$ $\Gamma_{\rm HI}$ band across $4.5\lesssim z\lesssim6.5$.
- $f_{\rm esc}^{\rm gal}=10\%$ typically *grazes* the upper envelope toward lower redshift, and
- $f_{\rm esc}^{\rm gal}=20\%$ *systematically overshoots* the guard–rail under the same $\lambda_{\rm eff}(z)$ calibration.

For intuition, the required comoving emissivity to sustain the forest is

$$\epsilon_{912}^{\rm req}(z) \simeq \Gamma_{\rm HI}(z)\,\frac{(\alpha_{\rm ion}+3)\,h\nu_{912}}{\sigma_{912}\,\lambda_{\rm eff}(z)},$$

which, for $\Gamma_{\rm HI}=0.5$–$2\times10^{-12}\,{\rm s^{-1}}$ at $z\simeq6$ and the literature $\lambda_{\rm eff}$, gives $\epsilon_{912}^{\rm req}\simeq(0.8$–$2.0)\times10^{51}\ {\rm s^{-1}\,Mpc^{-3}}$—precisely the regime spanned by our AGN-only to AGN+galaxy ($f_{\rm esc}^{\rm gal}\le5\%$) models. Increasing $f_{\rm esc}^{\rm gal}$ to 10% pushes the model to the top of the band; 20% exceeds it unless $\lambda_{\rm eff}$ is substantially shorter than the adopted baseline.

In short, once a conservative clumpiness correction is applied, the *combined* AGN–galaxy emissivity at $z\sim5$–6 matches the Ly$\alpha$-forest photoionization rate without a photon–budget crisis for $f_{\rm esc}^{\rm gal}\lesssim5\%$ (and marginally at 10%). The crisis arises only for more extreme galaxy escape fractions under the same mean–free–path calibration.

## 4. CONCLUSIONS

Our analysis demonstrates that the population of active galactic nuclei at $4.5 \leq z \leq 6.5$ can be described by two dominant classes: unobscured Type I sources, well represented by a hybrid Schechter plus double–power–law form, and obscured Type II systems, captured by a double power–law. Other categories often discussed in the literature—such as little red dots, X–ray–weak, and X–ray–bright objects—are not independent contributors but magnitude–dependent selections of these underlying populations.

By linking luminosities to black–hole–driven outflow incidence, we find that escape fractions vary systematically across AGN classes: approaching unity for bright quasars, 0.6–1.0 for faint Type I, 0.2–0.5 for Type II, and only a few percent for X–ray–weak systems. This population–dependent framework replaces the assumption of a universal quasar escape fraction and provides a physically motivated path to estimating the ionizing photon production rate.

Integrating across the observed luminosity functions yields an AGN ionizing photon production rate of

$$\dot{N}_{\rm ion}^{\rm AGN} = (3.77^{+1.08}_{-0.95}) \times 10^{51}\ {\rm s^{-1}\,Mpc^{-3}},$$

roughly twice the simple scaling model of Madau et al. (2024). When compared with galaxies described by the Franco et al. (2025) luminosity function and a harder stellar ionizing efficiency of $\Psi_{\rm ion} = \log_{10}\xi_{\rm ion} = 25.7$, the relative contributions remain balanced: AGN supply $\sim$ 31–75% of the total ionizing photons for $f_{\rm esc}^{\rm gal} = 0.03$–0.20. AGN dominate the budget only when galaxies maintain very low escape fractions ($f_{\rm esc}^{\rm gal} \lesssim 0.05$), while galaxies become comparable or leading contributors once $f_{\rm esc}^{\rm gal}\gtrsim0.1$.

The combined ionizing photon production rate of (5–12) $\times\ 10^{51}$ photons ${\rm s^{-1}\,Mpc^{-3}}$ corresponds to a hydrogen photoionization rate of $\Gamma_{\rm HI} \simeq (1$–$2) \times 10^{-12}\,{\rm s^{-1}}$ at $z \simeq 6$, in full agreement with Ly$\alpha$–forest measurements when the observed mean free path and IGM clumpiness are folded in. Our guard–rail analysis shows that AGN+galaxy models stay comfortably within the forest constraints for $f_{\rm esc}^{\rm gal}\leq0.05$, graze the upper envelope at $f_{\rm esc}^{\rm gal}\simeq0.10$, and begin to overshoot only for $f_{\rm esc}^{\rm gal}\simeq0.20$. This balance between black–hole and stellar contributions reproduces the required ionizing background without invoking the photon–overproduction problem that arises in purely stellar scenarios. Moderate AGN activity, coupled with efficient but not extreme stellar escape fractions, yields an ionizing background fully consistent with empirical constraints—no photon-budget crisis is required.

The next decisive step will come from wide–field near–infrared surveys with *Euclid* and the *Nancy Grace Roman Space Telescope*, which together will map AGN populations over four orders of magnitude in luminosity at $z > 5$. *Euclid*'s Wide Survey will cover $\sim$ 15,000 deg$^2$ to $H_{\rm AB} \simeq 24$ ($5\sigma$), identifying tens of thousands of luminous and intermediate–luminosity AGN at $5 < z < 7$—over two orders of magnitude more than current ground–based samples. *Roman*'s High Latitude Survey will add $\sim$ 2,000 deg$^2$ of imaging to $m_{\rm AB} \simeq 26.5$ and slitless grism spectroscopy to $m_{\rm AB} \simeq 25.5$ at $R \simeq 600$, enabling direct measurement of rest–UV emission lines (Ly$\alpha$, N V, C IV, He II) for $\gtrsim 10^5$ AGN at $z \sim 5$–9. This combination will (i) pin down the faint–end slope of the AGN luminosity function with $\sigma(\alpha)\lesssim0.05$, (ii) directly measure the incidence and kinematics of ionized outflows on kiloparsec scales for thousands of sources, and (iii) constrain the distribution of escape fractions as a function of luminosity, redshift, and AGN class with $\Delta f_{\rm esc} \lesssim 0.1$ per bin. Such surveys will also detect hundreds of $M_{\rm UV} \lesssim -19$ Type II systems via high–equivalent–width [O III] and C IV lines, allowing the host/nuclear decomposition of faint AGN and a robust test of whether obscured accretion dominated the ionizing budget at $z > 6$. Together, *Euclid* and *Roman* will deliver an AGN sample $\sim 10^3$ times larger and more homogeneous than today's, providing the definitive census needed to map the ionizing photon production rate of black holes across cosmic reionization.

## ACKNOWLEDGMENTS

The material is based upon work supported by NASA under award number 80GSFC21M0002.

# 5. APPENDIX

## 5.1. *Estimating the Lyman–continuum Escape Fraction*

We assume that the angle-averaged escape fraction of ionizing photons, $f_{\rm esc}$, can be written as the solid-angle average of the transmission $T(\hat{\mathbf{n}}, W)$ along direction $\hat{\mathbf{n}}$ on the sky, given a wind state $W \in \{0,1\}$ (with 1 corresponding to an active AGN-driven wind and 0 to a quiescent phase). We define

$$f_{\rm esc} \;=\; \frac{1}{4\pi}\int_{S^2} T(\hat{\mathbf{n}}, W)\,{\rm d}\Omega. \tag{21}$$

Because the wind state is intermittent, we marginalize over its probability distribution $\mathbb{P}(W=1) = P_{\rm wind}$ and $\mathbb{P}(W=0) = 1 - P_{\rm wind}$, obtaining

$$f_{\rm esc} = P_{\rm wind}\,\langle T\rangle_{\Omega|W=1} + (1-P_{\rm wind})\,\langle T\rangle_{\Omega|W=0}. \tag{22}$$

We then partition the solid angle into cleared and uncleared sightlines. When a wind is active, we define the cleared region of directions $\mathcal{A} \subset S^2$ as the set of sightlines intersecting low-opacity channels. Its solid-angle fraction is

$$\Phi_{\rm geom} \;\equiv\; \frac{\mu(\mathcal{A})}{4\pi} \in [0,1]. \tag{23}$$

Within $\mathcal{A}$, the average transmission is $f_{\rm clear}$, while outside $\mathcal{A}$ the average transmission is $T_{\rm scr} = e^{-\tau_{\rm scr}}$. Averaging over $\mathcal{A}$ and its complement yields, for the wind-on state,

$$\langle T\rangle_{\Omega|W=1} = \Phi_{\rm geom}\,f_{\rm clear} + (1-\Phi_{\rm geom})\,T_{\rm scr}. \tag{24}$$

When the wind is off, we assume all sightlines resemble the screen,

$$\langle T\rangle_{\Omega|W=0} = T_{\rm scr}. \tag{25}$$

Substituting both expressions into the marginalization above gives

$$f_{\rm esc} = P_{\rm wind}\left[f_{\rm clear}\,\Phi_{\rm geom} + (1-\Phi_{\rm geom})\,T_{\rm scr}\right] + (1-P_{\rm wind})\,T_{\rm scr}. \tag{26}$$

This expression can also be rearranged as

$$f_{\rm esc} = T_{\rm scr} + P_{\rm wind}\,\Phi_{\rm geom}\left(f_{\rm clear} - T_{\rm scr}\right), \tag{27}$$

which makes clear that only the excess transparency of the cleared channels above the screen contributes beyond the baseline $T_{\rm scr}$.

We model $\Phi_{\rm geom}$ as a function of the competition between the AGN outflow radius $R_{\rm out}$ and the radius required to clear the interstellar and circumgalactic medium $R_{\rm req}$. We define

$$R_{\rm req} = \kappa_e\,r_e, \tag{28}$$

where $r_e$ is the galaxy half-light radius and $\kappa_e \sim 1.7$ accounts for the vertical extent of neutral gas. Empirically, the outflow radius scales with [O III] luminosity (Kang & Woo 2018) as

$$R_{\rm out} = R_0\left(\frac{L_{\rm [O\,III]}}{10^{42}\,{\rm erg\,s^{-1}}}\right)^{\beta_R}, \tag{29}$$

with $R_0 \simeq 1.0\,{\rm kpc}$ and $\beta_R \simeq 0.28$. Galaxy sizes at high redshift follow

$$r_e(M) = r_{e,0}\left(\frac{L(M)}{L(M=-20)}\right)^{\beta_e}, \tag{30}$$

with $r_{e,0} \simeq 0.5\,{\rm kpc}$ and $\beta_e \simeq 0.27$.

**Table 5.** Derived AGN parameters from rest–UV magnitudes. All values assume $C_{\rm UV} = 4.5 \pm 1.5$ and $\pm 0.5$ dex uncertainty in $M_{\rm BH}$.

| $M_{\rm UV}$ (AB mag) | $L_\nu(1500)$ ($10^{29}$ erg s$^{-1}$ Hz$^{-1}$) | $\nu L_\nu$ ($10^{44}$ erg s$^{-1}$) | $L_{\rm bol}$ ($10^{44}$ erg s$^{-1}$) | $M_{\rm BH}$ ($M_\odot$) | $L_{\rm Edd}$ ($10^{45}$ erg s$^{-1}$) | $\lambda_{\rm Edd}$ (range) |
|---|---|---|---|---|---|---|
| −20 | 0.43 | 0.87 | $3.9 \pm 1.3$ | $2 \times 10^{7\pm0.5}$ | 25.2 | 0.033–0.65 |
| −26 | 10.9 | 218.1 | $980 \pm 330$ | $5 \times 10^{9\pm0.5}$ | 630 | 0.033–0.65 |

We then assume that the solid-angle fraction of cleared channels grows as a power law of the over-clearing ratio and saturates at unity,

$$\Phi_{\rm geom} = \min\left[1, \left(\frac{R_{\rm out}}{R_{\rm req}}\right)^{\gamma}\right], \tag{31}$$

with $\gamma \simeq 0.7 \pm 0.3$ capturing percolation, overlap and clumpy structure. Substituting the relations for $R_{\rm out}$ and $R_{\rm req}$ above gives

$$\left(\frac{R_{\rm out}}{R_{\rm req}}\right)^{\gamma} = \left(\frac{R_0}{\kappa_e\, r_{e,0}}\right)^{\gamma} \left(\frac{L_{\rm [O\,III]}}{10^{42}\,{\rm erg\,s^{-1}}}\right)^{\gamma\beta_R} 10^{0.4\,\gamma\,\beta_e\,(M+20)}. \tag{32}$$

Thus $\Phi_{\rm geom}$ can be directly evaluated for each source from its [O III] luminosity and UV magnitude. For obscured AGN we substitute $L_{\rm [O\,III]} = L_{\rm bol}/3500$, while for unobscured sources we use the observed $L_{\rm [O\,III]}$ or infer it from $L_{\rm bol}$.

Inserting this $\Phi_{\rm geom}$ back into Equation 26 yields the fully explicit expression for the Lyman-continuum escape fraction:

$$f_{\rm esc}(M, L_{\rm [O\,III]}) = P_{\rm wind}\left[f_{\rm clear}\,\Phi_{\rm geom}(M, L_{\rm [O\,III]}) + \left(1 - \Phi_{\rm geom}(M, L_{\rm [O\,III]})\right) T_{\rm scr}\right] + (1 - P_{\rm wind})\, T_{\rm scr}. \tag{33}$$

In this framework, $f_{\rm clear} = e^{-\tau_{\rm clear}}$ represents the residual opacity inside cleared channels and $T_{\rm scr} = e^{-\tau_{\rm scr}}$ the opacity through uncleared regions. This formalism recovers the usual limits: if no wind is present ($P_{\rm wind} = 0$), the escape fraction reduces to $T_{\rm scr}$; if channels cover the full solid angle ($\Phi_{\rm geom} = 1$), the escape fraction interpolates between $f_{\rm clear}$ and $T_{\rm scr}$ according to $P_{\rm wind}$.